\documentclass[a4paper,aps,prl,showpacs,twocolumn,nofootinbib]{revtex4}
\usepackage{amsmath,amssymb,graphicx}
\usepackage{color}

\newcommand{\be}{\begin{equation}}
\newcommand{\ee}{\end{equation}}
\newcommand{\bea}{\begin{eqnarray}}
\newcommand{\eea}{\end{eqnarray}}
\newcommand{\bean}{\begin{eqnarray*}}
\newcommand{\eean}{\end{eqnarray*}}

\newcommand{\de}{\delta}

\newcommand{\Ga}{\Gamma}

\def\id{{\rm 1\kern -2.5pt I}}

\begin{document}

\title{Can primordial magnetic fields be the origin of the BICEP2 data?}

\author{Camille Bonvin$^1$, Ruth Durrer$^2$ and Roy Maartens$^{3,4}$}

\affiliation{~\\$^1$Kavli Institute for Cosmology Cambridge and Institute of Astronomy, Madingley Road, Cambridge CB3 0HA, U.K.,
and\\
DAMTP, Centre
for Mathematical Sciences, Wilberforce Road, Cambridge CB3 0WA, U.K.\\
$^{2}$Universit\'e~de~Gen\`eve,~D\'epartement~de~Physique~Th\'eorique~and~CAP,\\
24 quai Ernest-Ansermet,~CH-1211~Gen\`eve,~Switzerland\\
$^{3}$Physics Department, University of the Western Cape, Cape Town 7535, South Africa\\
$^4$Institute of Cosmology \& Gravitation, University of Portsmouth, Portsmouth PO1 3FX, U.K.}

\date{\today}

\begin{abstract} 
If the 
B-mode signal in the CMB polarization seen by the \textsc{Bicep2} experiment is confirmed, it has dramatic implications for models of inflation. The result is also in tension with Planck limits on standard inflationary models.  It is therefore important to investigate whether this signal can arise from alternative sources. If so, this could lessen the pressure on inflationary models and the tension with Planck data. We investigate whether vector and tensor modes from primordial magnetic fields can explain the signal. We find that in principle, magnetic fields generated during inflation can indeed produce the required B-mode, for a suitable range of energy scales of inflation. In this case, the primordial gravitational wave amplitude is negligible, so that there is no tension with Planck and no problems posed for current inflationary models. However, the simplest magnetic model is in tension with Planck limits on non-Gaussianity in the trispectrum. It may be possible to fine-tune the magnetogenesis model so that this non-Gaussianity is suppressed. Alternatively, a weaker magnetic field can pass the non-Gaussianity constraints and allow the primordial tensor mode to be reduced to $r\simeq0.09$, thus removing the tension with Planck data and alleviating  the problems with simple inflationary models.

\end{abstract}


\maketitle

\section{Introduction}
Gravitational waves which can be observed in the polarization pattern of the cosmic microwave background (CMB) are often called the `holy grail' of inflation. Recently their experimental detection has been announced by the \textsc{Bicep2} collaboration~\cite{Ade:2014xna}.

This result, if confirmed by subsequent experiments, will be among the most important in cosmology since the discovery of the CMB. The reported tensor to scalar ratio of $r\simeq 0.2$ is very high. Such a high $r$ would  allow for a detailed study of the primordial tensor spectrum.  It would also imply an inflationary energy scale of $\simeq 1.4\times10^{16}\,$GeV, about 12 orders of magnitude above the highest energies reached by the LHC (Large Hadron Collider) at CERN. Furthermore, within the scenario of inflation, these gravitational waves are produced by the amplification of quantum fluctuations of the gravitational field itself. The result would therefore be evidence that the metric is a quantum field, i.e. our first observational indication of quantum gravity, even if only at the linear level.

The significance of this result demands rigorous scrutiny at the experimental level, extending the excellent work of the \textsc{Bicep2} collaboration. For simple inflationary models, the result is in tension with Planck data~\cite{Ade:2013zuv}, which require $r\lesssim0.11$. While experimental scrutiny proceeds, we require also a rigorous scrutiny at the theoretical level. One line of investigation is to revisit the simple inflationary models (see e.g.~\cite{Byrnes:2014xua}). Here we tackle another question -- are there alternative explanations of the signal? 

One of the first ideas that comes to mind is: could this signal arise from vector modes in the gravitational field which are not generated during inflation, but later in the evolution of the Universe by some inhomogeneous source? Vector modes are a potentially ideal source for B-polarization in the CMB, since their transfer function to B-modes is nearly 10 times larger than the one from tensor modes (see, e.g.~\cite{book}, Fig.~5.7).  Topological defects provide a potential origin of vector modes.  The possible contribution from defects to the \textsc{Bicep2} signal has been investigated by~\cite{Lizarraga:2014eaa} and it is found that defects cannot generate the observed signal~\footnote{{except if the effective inter-string distance is extremely large}}, but a small contribution from defects can alleviate the tension with the Planck results.

Here we investigate another possibility, primordial magnetic fields. Magnetic fields which are generated causally after inflation have blue spectra, $n_B=2$, and cannot leave an observable imprint in the CMB~\cite{spec}. However, magnetic fields can also be generated during inflation by couplings of the electromagnetic field to the inflaton or to the metric~\cite{TW}.
In this case, they can have an arbitrary spectrum with $n_B>-3$, where $n_B\simeq -3$ is scale invariant. The imprint of such magnetic fields in the CMB has been studied extensively; see e.g.~\cite{pedro} and references therein. It has been found that all modes (scalar, vector and tensor) contribute with similar amplitudes to the CMB temperature anisotropies and to the E-polarization, but the B-polarization of the compensated mode is dominated by the vector mode~\cite{Shaw:2009nf}.

We show below that the combination of  compensated and passive modes from  a magnetic field alone can reproduce very well the \textsc{Bicep2} result without invoking primordial tensor modes (i.e. taking $r$ to be negligible). However,  
if we require that the initial magnetic field be Gaussian, then limits on non-Gaussianity are in tension with a pure magnetic field solution. 
On the other hand, the inclusion of a weaker magnetic field, consistent with non-Gaussianity bounds, can reduce the required inflationary tensor contribution to $r\simeq 0.09$, thus removing the present tension with  temperature data from Planck. 

\section{Magnetic modes and B-polarization}

We consider a magnetic field generated during inflation, which is nearly scale-invariant  $n_B=-2.9$. Such a field produces both `compensated' and `passive' modes. The first type is an isocurvature mode, compensated by free-streaming neutrinos after they  decouple, and it is sourced until late times. The passive mode, which is adiabatic, is no longer sourced after neutrino decoupling, and its amplitude depends logarithmically on the scale of generation~\cite{Shaw:2009nf,CC}. The inflationary and magnetic passive modes in general have a higher amplitude than the magnetic compensated mode. A third `acausal' (inflationary) magnetic mode discovered in~\cite{CCD} is always scale-invariant and, for $n_B=-2.9$, has the same characteristics as the passive mode. We therefore do not discuss it separately in the following -- it can simply be added to the passive mode, enhancing the pre-factor $F$ in eq.~\eqref{defF} (see also footnote 3).

The passive magnetic mode mimics an inflationary spectrum with scalar spectral index $n_S=2n_B+7$ and tensor index $n_T=2n_B+6$. It can only be distinguished from an inflationary spectrum via higher order correlators (bi- and trispectrum). We can characterize 
the passive mode by its curvature perturbation,
\bea
\hspace*{-0.2cm}P_{\zeta_B}(k) &=& A_p(k/k_*)^{2n_B+6},\\ 
A_p &=&1.87\times 10^{-13}\frac{(0.05)^{2n_B+6}}{\Ga^2(n_B/2+3/2)}F\left[\frac{B_1}{1\,\rm{nG}}\right]^4\!\!\!, \label{defF}\\
F&=&[\log(T_*/T_\nu)+1/2]^2 , \label{e:Fpassive}
\eea
and the compensated mode by the density fluctuation which it induces:
\bea
P_{\de_B} &=& A_c(k/k_*)^{2n_B+6},\\
 A_c &=& 1.71\times 10^{-13}\frac{(0.05)^{2n_B+6}}{\Ga^2(n_B/2+3/2)}\left[\frac{B_1}{1\,\rm{nG}}\right]^4\!\!,
\eea
where $T_*$ is the energy scale at which the field started to evolve freely in the radiation dominated era, e.g. after reheating, and $T_\nu\simeq \,$MeV is the scale of neutrino decoupling. More details are found in~\cite{Shaw:2009nf,CC}.

We consider three magnetic cases:
\begin{eqnarray}
{\rm (M1)}&&B_1=1.83\,{\rm nG}, ~~T_*=10^{14}\,{\rm GeV}, \\
{\rm (M2)}&&B_1=3.04\,{\rm nG}, ~~T_*=10^3\,{\rm GeV},\\
{\rm (M3)} &&B_1=5.5\,{\rm nG}, ~~T_*=57\,{\rm MeV}.
\end{eqnarray}
Here $B_1$ is the amplitude of the magnetic field today at the scale 1\,Mpc. The case (M3) has been included as illustration with a large contribution from the compensated mode, and has a non-realistic value of $T_*$. 

\begin{figure}[!htbp]
\centering
\includegraphics[width=0.49\textwidth]{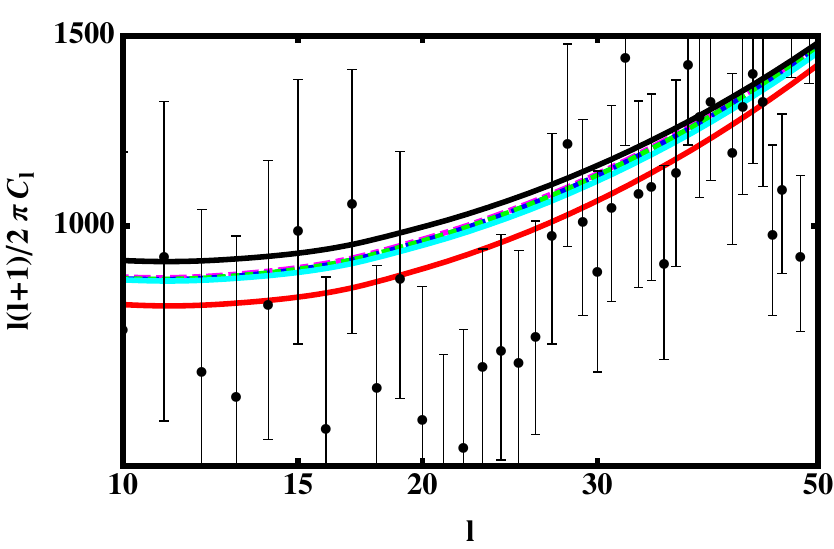}
\caption{Temperature angular auto-correlation spectra: from an inflationary scalar mode (red, bottom solid line), inflationary scalar + tensor modes with $r=0.11$ (cyan, middle solid line),  inflationary scalar + tensor modes with $r=0.2$ (black, top solid line) and from an inflationary scalar mode + magnetic modes, in case (M1) (green, short-dashed), (M2) (blue, long-dashed) and (M3) (magenta, dot-dashed). Data points are from Planck.
\label{ctt}
}
\end{figure}

In  Fig.~\ref{ctt} we compare  the contribution of a primordial magnetic field to the temperature spectra (calculated with the modified CAMB code of~\cite{Shaw:2009nf}) with the contribution of a primordial tensor mode with $r=0.11$ and with $r=0.2$ at the pivot scale $k_\lambda=0.002\,{\rm Mpc}^{-1}$. The three magnetic models (M1-M3) make a contribution which is just at the edge of what is allowed by Planck temperature measurements, $r<0.11$, and well below the contribution from a primordial tensor with $r=0.2$.\footnote{Note that while (M1) and (M2) are completely compatible with Planck at high multipoles, (M3) generates a too large contribution from the compensated vector mode.}

\begin{figure}
\includegraphics[width=0.49\textwidth]{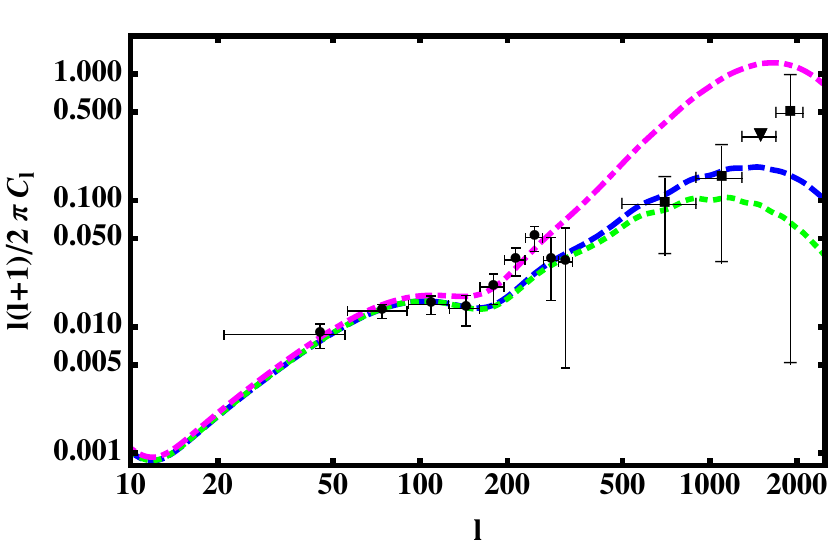} \\
\caption{\label{f:pureBbmode} The B-mode polarization spectrum, i.e. the sum of the scalar lensing B-mode with the magnetic modes in cases (M1), (M2) and (M3) (as in Fig.~\ref{ctt}). Data points are from \textsc{Bicep2} (circles) and \textsc{Polarbear} (squares).}
\end{figure}

Figure~\ref{f:pureBbmode} shows a fit to the \textsc{Bicep2} and \textsc{Polarbear} data~\cite{polarbear} from pure magnetic field B-modes plus a scalar lensing B-mode. It is interesting that magnetic fields with $n_B>-3$, which is required in order to evade an infrared divergence, automatically lead to a blue tensor spectrum which seems to be favored by the data~\cite{blueT}. The value of $n_B=-2.9$ adopted here yields $n_T=0.2$. 
In all three cases, the dominant contributions to the magnetic field B-modes are from the passive tensor mode and the compensated vector mode. A large value of $T_*$ (M1) amplifies the passive tensor mode with respect to the vector mode so that the magnetic contribution is essentially indistinguishable from the inflationary prediction at all multipoles. On the other hand, for smaller values of $T_*$ (M2 and especially M3), the contribution from the vector mode is enhanced, leading to an increase of the signal at high multipoles. 
The plots show that magnetic fields (M1) and (M2) can mimic the $r\simeq0.2$ inflationary prediction very well and are also compatible with the \textsc{Polarbear} data at sub-degree scales. Moreover, thanks to the slightly blue spectrum of the magnetic passive tensor mode $n_T=0.2$ and to the fact that the magnetic compensated vector mode contributes in a negligible way to the temperature, the magnetic field contribution is compatible with the temperature spectrum measured by Planck. Note that a bluer magnetic field with e.g. $n_B=-2.8$ would still fit the polarisation data well, while reducing even more the contribution to the temperature spectrum. We defer a full  MCMC analysis of the magnetic field parameters to subsequent work.

 {\em In principle, the \textsc{Bicep2} data can be explained via magnetic fields, with inflation producing negligible tensor modes and thus avoiding tension with Planck and problems for inflationary model-building. } 

\section{Constraints from non-Gaussianity}
\label{sec:NG}

If the magnetic field distribution is Gaussian, its energy momentum tensor is the square of a Gaussian and its non-Gaussianity is mainly of the local type. The Planck constraints~\cite{PlanckGauss} on the bispectrum then imply $B_1\lesssim 2-3\,$nG~\cite{Caprini:2009vk}.  The amplitude $B_1$ (or the pre-factor $F$) needed to reproduce the B-polarization observed by \textsc{Bicep2} is just small enough not to spoil the bispectrum constraint from Planck. On the other hand, stronger constraints have recently been shown to arise from the trispectrum~\cite{Trivedi:2013wqa}. In particular, the passive scalar mode leads to the strongest constraint: $B_1\lesssim 0.9\,$nG for $n_B=-2.8$  and $T_*=10^{14}\,$GeV, which corresponds to $B_1\lesssim 1.2\,$nG for $n_B=-2.9$ (see Eq. (48) in~\cite{Trivedi:2013wqa}). The amplitude of the magnetic field needed to reproduce the \textsc{Bicep2} data generates therefore a slightly too large trispectrum in the CMB, if one assumes that the magnetic field itself is Gaussian (the above constraints assume Gaussianity of the magnetic field). Note however that the magnetic field contribution to the trispectrum has been calculated in~\cite{Trivedi:2013wqa} for a limited number of shapes only and that a full calculation may slightly soften the 1.2\,nG bound due to possible cancellations between the different shapes.
\footnote{Note also that the constraint $B_1\lesssim 0.05\,$nG coming from the trispectrum of the acausal magnetic mode~\cite{Trivedi:2013wqa}, is not in disagreement with our result. The contribution of the acausal magnetic mode to both the B-polarization and the trispectrum is sensitive to the combination $B_1^4F$. As shown in~\cite{CCD} if $B_1$ is not amplified during reheating, $F$ is very large for the acausal mode, $F_{\rm acaus}\sim 10^5 F_{\rm passive}$, (where $F_{\rm passive}$ is the value of $F$ given in Eq.~(\ref{e:Fpassive})) meaning that a correspondingly smaller $B_1$ is sufficient to fit the \textsc{Bicep2} data. Our result therefore still holds if we include the acausal magnetic mode: the particular values of $F$ and $B_1$ will be different but the effect on the B-polarization (and on the trispectrum)  remains the same. 
Note however that one difference arises: since the amplitude $B_1$ needed to fit the \textsc{Bicep2} data is strongly reduced, the effect of the compensated mode (which depends {\it only} on $B_1$ and not on the combination $B_1^4F$) becomes completely negligible, and of course the constraint on $B_1$ becomes much stronger as discussed in~\cite{Trivedi:2013wqa}.}

\begin{figure}
\includegraphics[width=0.49\textwidth]{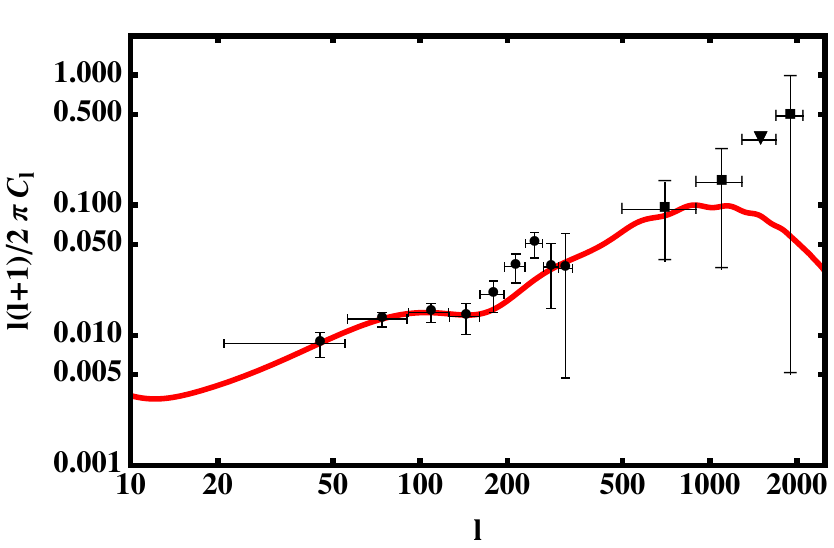} \\
\caption{\label{f:Bmix} The B-mode from lensed scalars plus a primordial tensor spectrum with $r=0.09$ at $k_\lambda=0.002\,\rm{Mpc}^{-1}$, a magnetic field with $n_B=-2.9$ and $B_1=1.2\,$nG, and a dust contribution of $0.0025\,(\mu{\rm K})^2$. The sum of the primordial and the magnetic contribution reproduce effectively the $r=0.16$ cited in~\cite{Ade:2014xna} after dust removal.
}
\end{figure}

To evade the constraint $B_1\lesssim 1.2\,$nG, we could try to build inflationary magnetogenesis models with non-Gaussian magnetic fields whose trispectra are suppressed with respect to the Gaussian case. Even though a logical possibility, this seems to be an artificial and unnatural alternative to the high-$r$ inflationary tensor modes. Instead, we can use a reduced contribution from magnetic fields that is consistent with non-Gaussianity bounds. This will take some pressure off the inflationary tensor modes, reducing the $r$ value needed to match the data. In Fig.~\ref{f:Bmix} we give an example to show that $r=0.09$ can be achieved by adding magnetic fields whose bi- and trispectra are consistent with current bounds and including a small amount of dust, $[\ell(\ell+1)C_\ell^{BB}/2\pi]_{\rm dust} \simeq 0.0025\,(\mu{\rm K})^2$. Note that this combination of a primordial tensor mode and a magnetic mode also respects the temperature Planck bounds of $r<0.11$.

\section{Conclusions\label{s:con}}

B-modes from magnetic fields can reproduce the \textsc{Bicep2} results with no contribution from inflationary gravitational waves, i.e. with $r\simeq0$. This requires, however, that the fields are generated during inflation with non-Gaussian statistics, in such a way that their energy-momentum tensor is nearly Gaussian. As far as we are aware, no specific mechanism to produce such fields has been proposed in the literature so far.

If Gaussian magnetic fields are generated during inflation, then the non-Gaussianity induced by the fields that are required to replace the $r\simeq 0.2$ tensor mode, are in tension with the trispectrum limits from Planck~\cite{Trivedi:2013wqa}. 

Nevertheless, a reduced magnetic contribution together with a small amount of dust can bring the required tensor amplitude down to $r\simeq 0.09$.  This mitigates the tension with Planck bounds~\cite{Ade:2013zuv}, which require a relatively strong negative running of the scalar spectrum if $r\simeq 0.2$.

Finally, we note that there is a strong-coupling problem of magnetogenesis. If a magnetic field is generated during inflation by a term ${\cal L} \supset f^2(\phi)F_{\mu\nu}F^{\mu\nu}$ in the Lagrangian, the `running' of $f$ during inflation, which is needed for a nearly scale-invariant spectrum, requires that the coupling of the electromagnetic field to the electron $\propto e/f(\phi)$ has been strong during most of the inflationary epoch and perturbation theory cannot be trusted~\cite{Demozzi}. This can be alleviated by postulating a very low inflationary scale, or a very blue magnetic field spectrum~\cite{Raj1}.  It has been shown~\cite{Raj2} that for an inflationary scale $\sim10^{16}\,$GeV, $f$ is so severely constrained that only fields with $B_1\lesssim 10^{-30}\,$G can be generated.  However, there are also other ways to evade the strong-coupling problem, for example breaking gauge-invariance during inflation~\cite{CCD} or choosing different couplings to the electromagnetic field.

Taking these caveats into account, we believe it is fair to say that in principle, the observed B-mode signal could be due to inflationary magnetic fields. However, even though contrary to defects~\cite{Lizarraga:2014eaa}, B-modes from magnetic fields have the right shape, the simplest models are problematic since they are in tension with the upper limit from Planck on the trispectrum. On the other hand, a weaker magnetic field compatible with non-Gaussianity constraints can reproduce the observed B-mode, while reducing the primordial signal to $r\simeq 0.09$, thus removing the tension with Planck data.\vspace*{-0.6cm}

\[\]{\bf Acknowledgements:}
It is a pleasure to thank Chiara Caprini, Anthony Challinor, Dani Figueroa, Rajeev Jain, Andrii Neronov, Christophe Ringeval, Lorenzo Sorbo and Kandaswamy Subramanian for discussions and Richard Shaw for his help with the modified CAMB code.
We also thank Brian G. Keating for calling to our attention the \textsc{Polarbear} data.
CB is supported by King's College Cambridge. RD acknowledges support from the Swiss National Science Foundation.
RM is supported by the South Africa Square Kilometre Array Project,  the South African National Research Foundation and the UK Science \& Technology Facilities Council (grant ST/K0090X/1). 
\end{document}